\documentclass[a4paper]{jpconf}
\usepackage{graphicx}
\begin{document}
\title{Can CCM law properly represent all extinction curves?}

\author{Anna Geminale $^{1,2}$, Piotr Popowski $^{2}$}

\address{$^{1}$Department of Astronomy, Vicolo dell'Osservatorio 3, 35122
Padova, Italy}
\address{$^{2}$ Max Planck Institute for Astrophysics,
Karl-Schwarzschild-Str. 1, Postfach 1317, D-85741 Garching, Germany}

\ead{geminale@pd.astro.it, popowski@mpa-garching.mpg.de}

\begin{abstract}
We present the analysis of a large sample of lines of sight with 
extinction curves covering wavelength range from near-infrared (NIR)
to ultraviolet (UV). We derive total to selective extinction ratios
based on the Cardelli, Clayton \& Mathis (1989, CCM) law, which is
typically used to fit the extinction data both for diffuse and dense 
interstellar medium. We conclude that the CCM law is able to fit most 
of the extinction curves in our sample. We divide the remaining lines
of sight with peculiar extinction into two groups according to two
main behaviors: a) the optical/IR or/and UV wavelength region cannot be
reproduced by the CCM formula; b) the optical/NIR and UV extinction
data are best fit by the CCM law with different values of $R_V$. We
present examples of such curves. The study of both types of peculiar
cases can help us to learn about the physical processes that affect
dust in the interstellar medium, e.g., formation of mantles on the
surface of grains, evaporation, growing or shattering.
\end{abstract}

\section{Introduction}

Interstellar grains affect starlight which passes through them by 
absorbing and scattering photons. These two physical processes
produce the interstellar extinction which depends on the properties of
dust grains, e.g., size distribution and composition. The
description of the average extinction curve of our Galaxy as a
function of the wavelength from the infrared to ultraviolet can be
found in Savage \cite{Savage79}. Extinction curve shows some evident
features: it rises in the infrared, it shows a slight knee in the
optical, it is characterized by a bump at 2175$\AA$, and it rises in
the far-ultraviolet. These features are common between different
environments. The properties of interstellar grains are different in
diffuse and dense interstellar medium and thus also the extinction
changes. CCM \cite{Cardelli89} found an average extinction law valid
over the wavelength range $0.125\mu m \leq \lambda \leq 3.5\mu m$,
which is applicable to both diffuse and dense regions of the
interstellar medium. This extinction law depends on only one parameter
$R_V=A_V/E(B-V)$. The $R_V$ parameter ranges from about 2.0 to about
5.5 (with a typical value of 3.1) when one goes from diffuse to dense 
interstellar medium and thus $R_V$ characterizes the region that
produces the extinction.

\begin{figure}
\includegraphics[height=9.5cm,width=8cm]{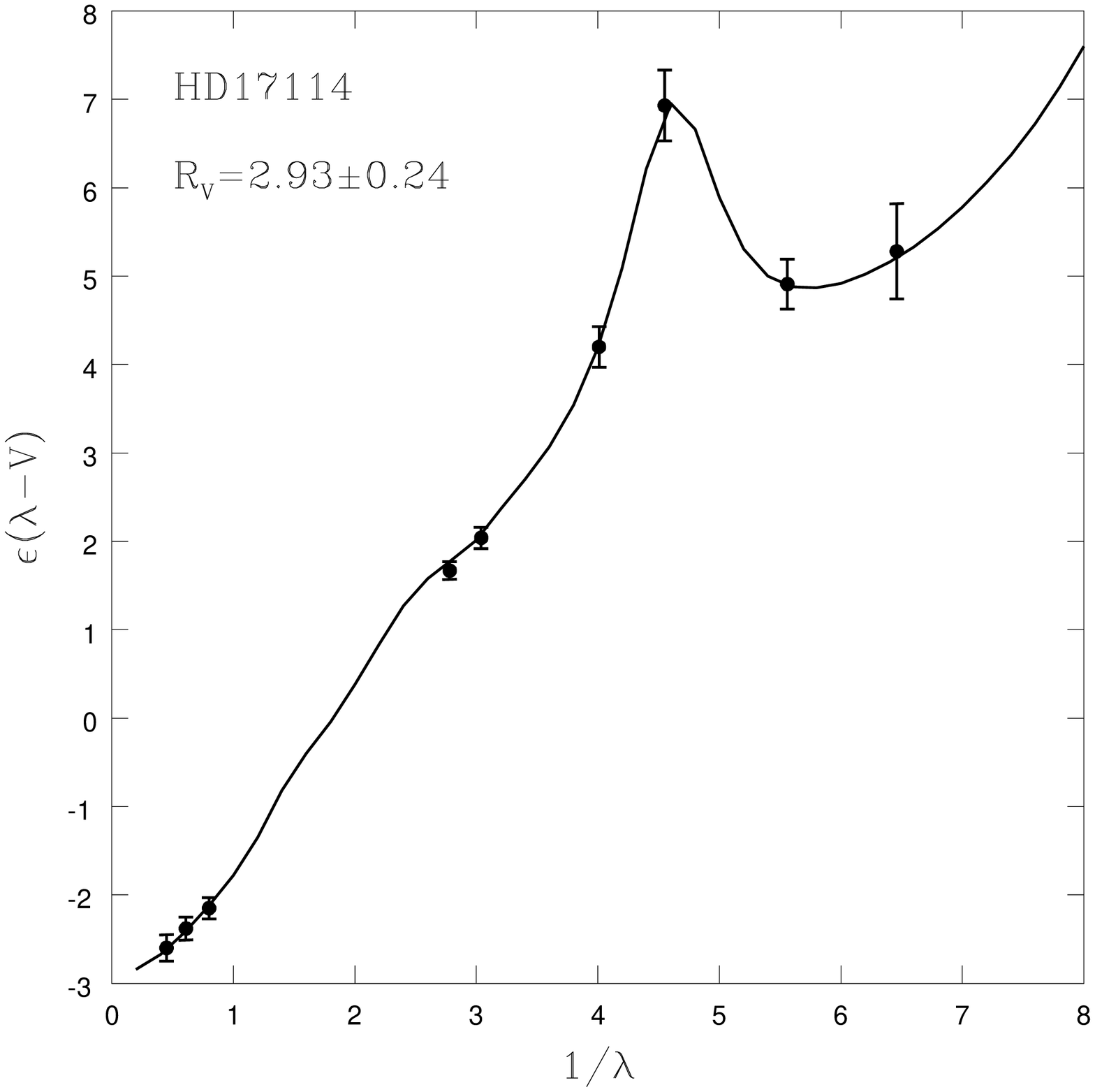}
\includegraphics[height=9.5cm,width=8cm]{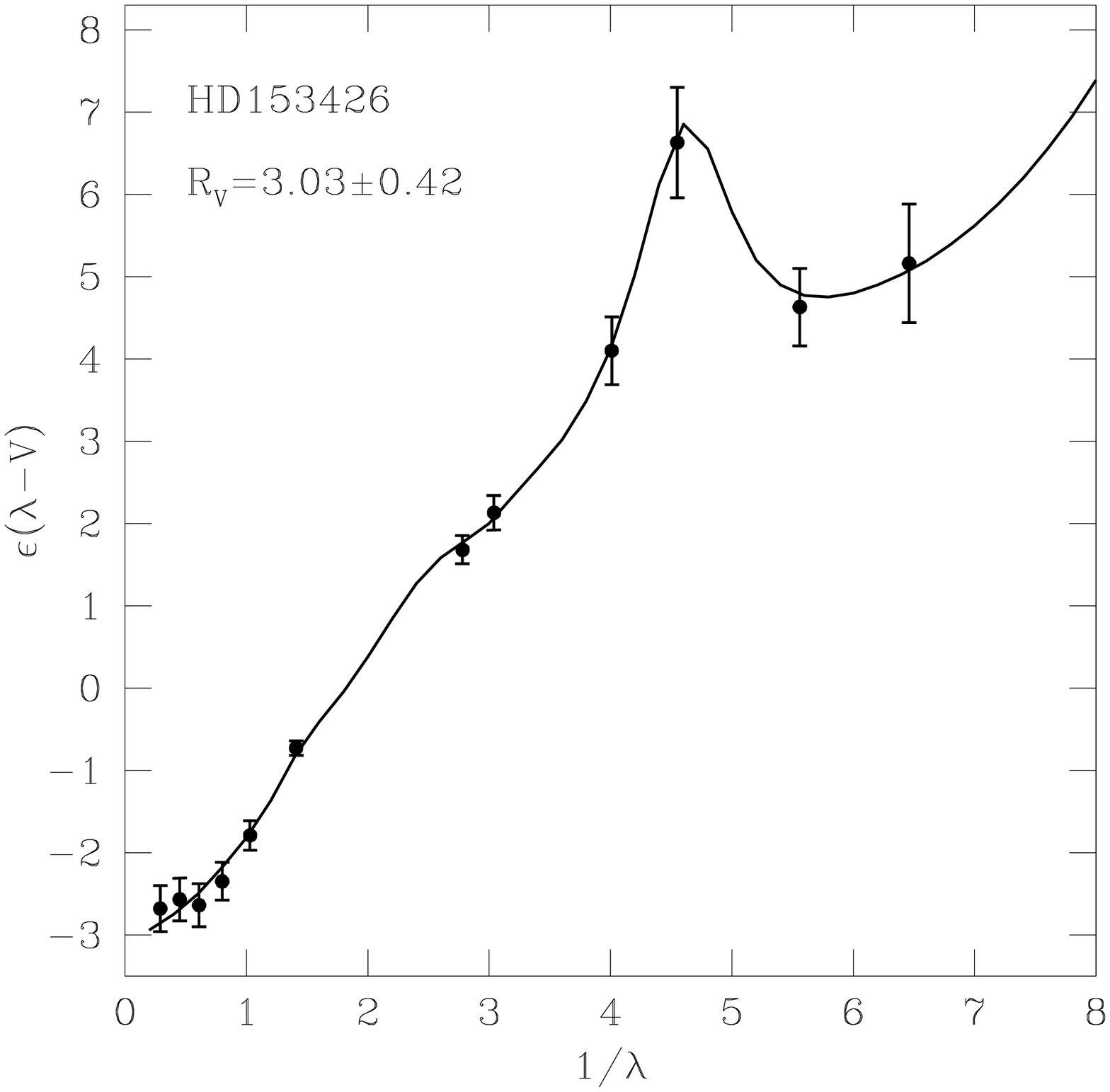}
\caption{Examples of extinction curves for which the CCM
law works in the entire spectral range from IR to UV.}
\label{fig1}
\end{figure}

If one knows the value of $R_V$ along a particular line of sight, one can
obtain the extinction curve from the infrared to ultraviolet using the
CCM law (see Figure \ref{fig1}):
\begin{equation}
\frac{A_{\lambda}}{A_V}=a(x)+b(x) \cdot R_V^{-1},
\label{CCMlaw}
\end{equation}
where $x=1/\lambda$, and $a(x)$ and $b(x)$ are the
wavelength-dependent coefficients.

There are different ways to obtain $R_V$ using the NIR or UV
extinction data. Wegner \cite{Wegner03} computed $R_V$ values for
the sample of 597 OB stars with known near-infrared magnitudes.
He assumed that, in the infrared spectral region, the 
normalized extinction curve is proportional to $\lambda^{-3}$ or 
$\lambda^{-4}$. Extrapolating the IR interstellar extinction curve to 
$1/\lambda=0$ he derived $R_V$ as:
\begin{equation}
R_V=- \left [ \frac{E(\lambda-V)}{E(B-V)} \right] _{\lambda
\longrightarrow \infty}
\label{RvWegner}
\end{equation}
Gnaci{\'n}ski \& Sikorski \cite{Gnacinski99} applied the $\chi^2$ 
minimization method to compute the $R_V$ values for a sample of
ultraviolet (UV) extinction data using the linear relation
(\ref{CCMlaw}). Geminale \& Popowski \cite{GP04} extended the analysis
from \cite{Gnacinski99} by using non-equal weights derived from
observational errors to determine $A_V$ and $R_V$ values toward a
sample of stars with known ultraviolet color excesses.

In this paper we use both optical/infrared ($URIJHKLM$) and UV 
extinction data to obtain $R_V$ values for
a sample of 436 lines of sight.  We arrive at two main conclusions:
(i) there are lines of sight with extinction in the optical/IR or/and
UV which generically don't follow the CCM law, and (ii) there are
lines of sight which show an extinction curve that cannot be
reproduced with a single $R_V$ value in the whole wavelength range.

\section{Theoretical Basis}

We normalize the extinction in a standard way: 
\begin{equation}
\epsilon(\lambda-V)=\frac{E(\lambda-V)}{E(B-V)}.
\end{equation}
The absolute extinction may be deduced from the
relative extinction by using a total-to-selective extinction ratio $R_V$:
\begin{equation}
R_V=\frac{A_V}{E(B-V)}.
\end{equation}
Then:
\begin{equation}
\epsilon(\lambda-V)=\frac{E(\lambda-V)}{E(B-V)}=
\frac{A_{\lambda}-A_{V}}{E(B-V)} 
= R_V \left \{\frac{A_{\lambda}}{A_{V}}-1 \right \}.
\label{Eps}
\end{equation}
For each individual band, equation (\ref{CCMlaw}) and (\ref{Eps}) can
be combined to derive an $R_V$ value. More generally, the $\chi^2$ 
minimization can be used to obtain the $R_V$ value that provides the
best CCM fit to all observed extinction data. Gnaci{\'n}ski \&
Sikorski \cite{Gnacinski99} suggested the following $\chi^2$, which
can be minimized to derive $R_V$:
\begin{equation}
\chi^2=\sum_{i=1}^{N_{\rm bands}} \left \{ E(\lambda_i-V)- E(B-V)
\cdot [R_V
(a(x_i)-1) + b(x_i)] \right \} ^2,
\end{equation}
where $a(x_i)$ and $b(x_i)$ are the coefficients of the CCM
law. Following our previous work \cite{GP04} we use an improved
weighted formula to find $R_V$. We minimize the following $\chi^2$:
\begin{equation}
\chi^2=\sum_{i=1}^{N_{\rm bands}} w_{\lambda_i} \{
\epsilon(\lambda_i-V) - [R_V(a(x_i) -1)+
b(x_i)] \} ^2 ~E^2(B-V)
\label{chiweighted}
\end{equation}
where $\omega_i \equiv 1/\sigma_i^2$ are the weights associated with
each band. 

Minimizing equation (\ref{chiweighted}) with respect to $R_V$, we find:
\begin{equation}
R_V=\frac{\sum_{i=1}^{N_{\rm bands}} \{ (a(x_i)-1)\cdot
(\epsilon(\lambda_i-V)-b(x_i))/ \sigma_i^2
\}}{\sum_{i=1}^{N_{\rm bands}} \{ (a(x_i)-1)^2/
\sigma_i^2 \} }
\label{Rvweighted}
\end{equation}
where here $\sigma_i$ values are taken from Wegner \cite{Wegner02} and
they were computed according to: 
\begin{equation}
\sigma_i^2 \equiv  \sigma^2 [\epsilon(\lambda_i-V)] =\left[
\frac{1}{E(B-V)} \right]^2
\, \left \{\sigma^2[E(B-V)] + [E(\lambda_i-V)]^2 \,
\sigma^2[E(\lambda_i-V)] \right \}
\label{sigma}
\end{equation}
We estimate the error in $R_V$ from:
\begin{equation}
\sigma(R_V) \equiv \sum_{j=1}^{N_{\rm bands}} \left
|\frac{\partial R_V}{\partial
\epsilon(\lambda_j-V)} \right| \cdot \sigma_j
= \frac{1}{\sum_{i=1}^{N_{\rm bands}}
(a(x_i)-1)^2/\sigma_i^2} \cdot \sum_{j=1}^{N_{\rm bands}} \left | 
\frac{a(x_j)-1}{\sigma_j} \right|
\label{errabs}
\end{equation}

\section{Data}
We use a sample of 436 lines of sight with optical/IR and UV
extinction data reported by Wegner \cite{Wegner02}. In Wegner's
computation the ultraviolet photometry is taken from
Wesselius et al. \cite{Wesselius82} and based on \emph{Astronomical 
Netherlands Satellite} (ANS); infrared magnitudes in $J,H,K,L,M$
passbands are originate mostly from the catalog of Gezari, Schmitz and
Mead \cite{Gezari84} and Gezari et al. \cite{Gezari93}. The $R$ and $I$
magnitudes with accuracy of $\pm 0.01$ mag are taken from Johnson 
\cite{Johnson66} and Fernie \cite{Fernie83}. The spectral
classification and $UBV$ data which have the accuracy of $\pm 0.01$
mag come from the SIMBAD database.

The effective wavelengths of the optical/IR bands are: $\lambda_U=0.36
\mu m$, $\lambda_R=0.71 \mu m$, $\lambda_I=0.97\mu m$, $\lambda_J=1.25
\mu m$, $\lambda_H=1.65 \mu m$, $\lambda_K=2.2 \mu m$, $\lambda_L=3.5
\mu m$, $\lambda_M=4.8 \mu m$; whereas the effective wavelengths of
the UV bands are: 0.1549, 0.1799, 0.2200, 0.2493, and 0.3294 $\mu m$.

From the original sample, we exclude 20 lines of sight because of
their negative value of $R_V$. Negative $R_V$ values are typically the
result of noisy data for the lines of sight with small $E(B-V)$. 

\section{Results}

\label{Results}

\subsection{$\chi^2$ test}

\begin{figure}
\includegraphics[height=9.5cm,width=8cm]{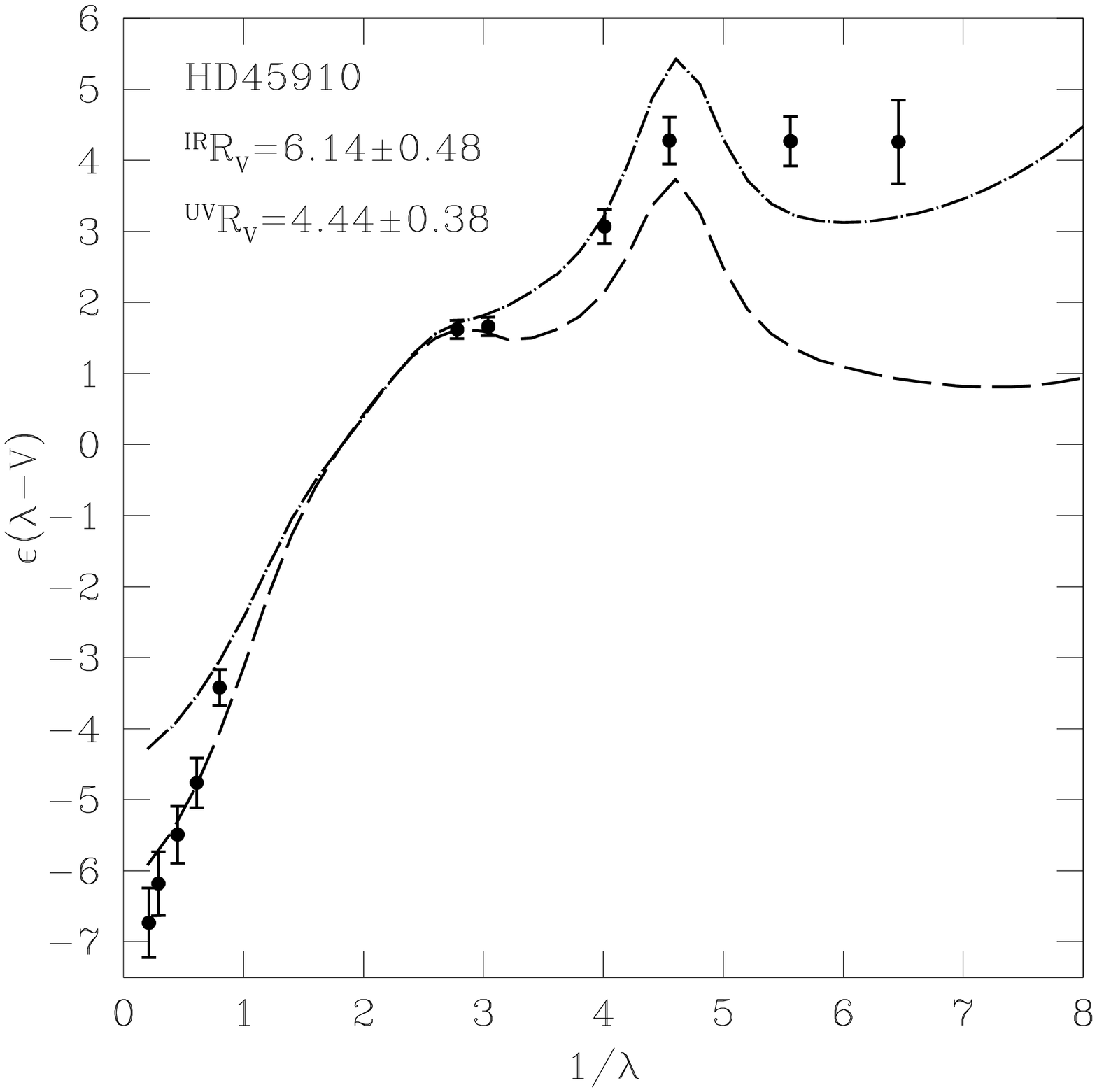}
\includegraphics[height=9.5cm,width=8cm]{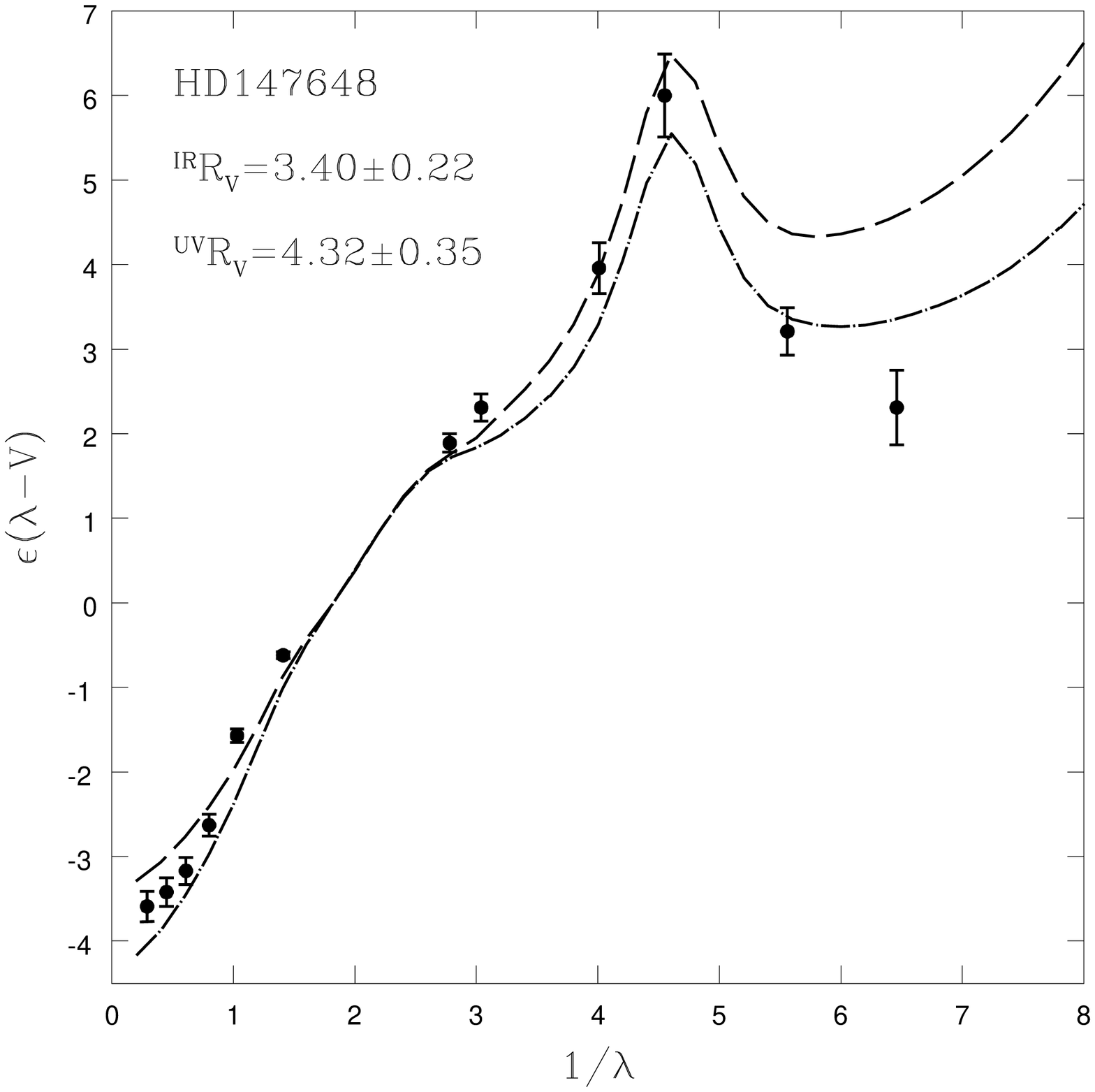}
\caption{Examples of extinction curves with $\chi^2/dof>2.0$. The long
dashed line represents the CCM curve
which is the best fit to the IR data and the point-short dashed line
represents the CCM curve which best follows the UV data.}
\label{fig2}
\end{figure}

A useful method to test if all extinction data points are well fitted
by the CCM law is to compute $\chi^2$ based on equation
(\ref{chiweighted}) normalized to the number of degrees of freedom,
which is equal to the number of observed points minus the number of fitted
parameters (in our case the only parameter is $R_V$). We note that our
average $\chi^2/dof$ is not equal to the expected value of 1; since
the errors are taken from Wegner
(2002) and we do not want to modify them, we do not renormalize our
errors requesting $<\chi^2/dof>=1$. Therefore it is safer to treat our
$\chi^2/dof$ as a measure of a relative rather than absolute quality
of the CCM fit in the optical/IR and UV wavelength ranges. We select
outliers considering the tail of our $\chi^2/dof$ distribution. We
assume that the values $\chi^2/dof>2.0$ are indicative of the lines of
sight with extinction curves not well fitted by the CCM law. We find
that 25\% of the lines of sight of our sample shows this disagreement
with the CCM law. Figure \ref{fig2} displays two extinction curves 
representative of this group.

\subsection{Test for the universality of CCM law}

We further analyze the lines of sight with $\chi^2/dof < 2.0$ for
which we expect the CCM law to fit well all observed extinction data
from optical/IR to UV. The usual assumption is that the knowledge of
the $R_V$ value obtained from the IR part of the extinction curve may
be used to obtain the entire extinction curve by using the CCM law. We
critically test this assumption using two sets of $R_V$ values for
each line of sight: the infrared $R_V$ ($^{\rm{IR}}\!R_{V}$) and
ultraviolet $R_V$ ($^{\rm{UV}}\!R_{V}$). We compute the following
statistic: 
\begin{equation}
\delta=
\frac{^{\rm{UV}}\!R_{V}-\, ^{\rm{IR}}\!R_{V}}{\sqrt{\sigma^2[^{\rm{UV}}\!R_{V}]+
\sigma^2[^{\rm{IR}}\!R_{V}]}}.
\label{deviation}
\end{equation}
Specifically, when $|\delta| \geq 2.0$ we classify the line of sight 
as anomalous in the sense that the CCM law is not able to
reproduce the whole extinction curve with a single value of $R_V$.

\begin{figure}
\begin{center}
\includegraphics[width=85mm]{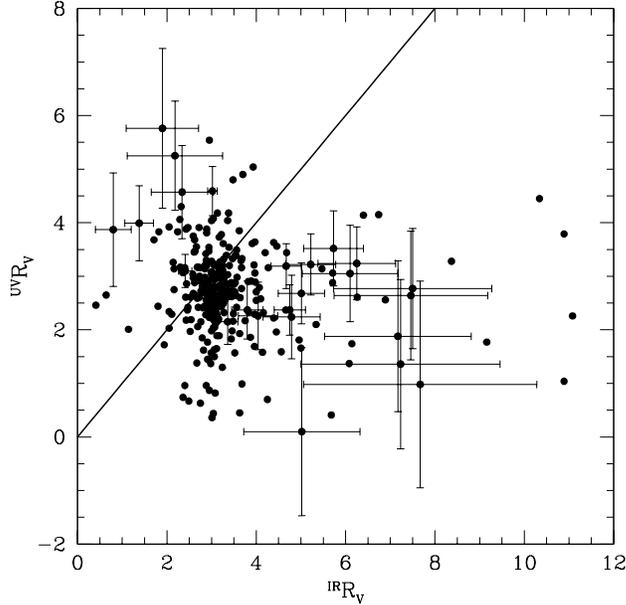}
\caption{The comparison between the $R_V$ values obtained from
optical/IR and UV extinction data. The points with error bars are
those for which $|\delta| \geq 2.0$.}
\label{fig3}
\end{center}
\end{figure}

\begin{figure}
\includegraphics[height=10cm,width=8cm]{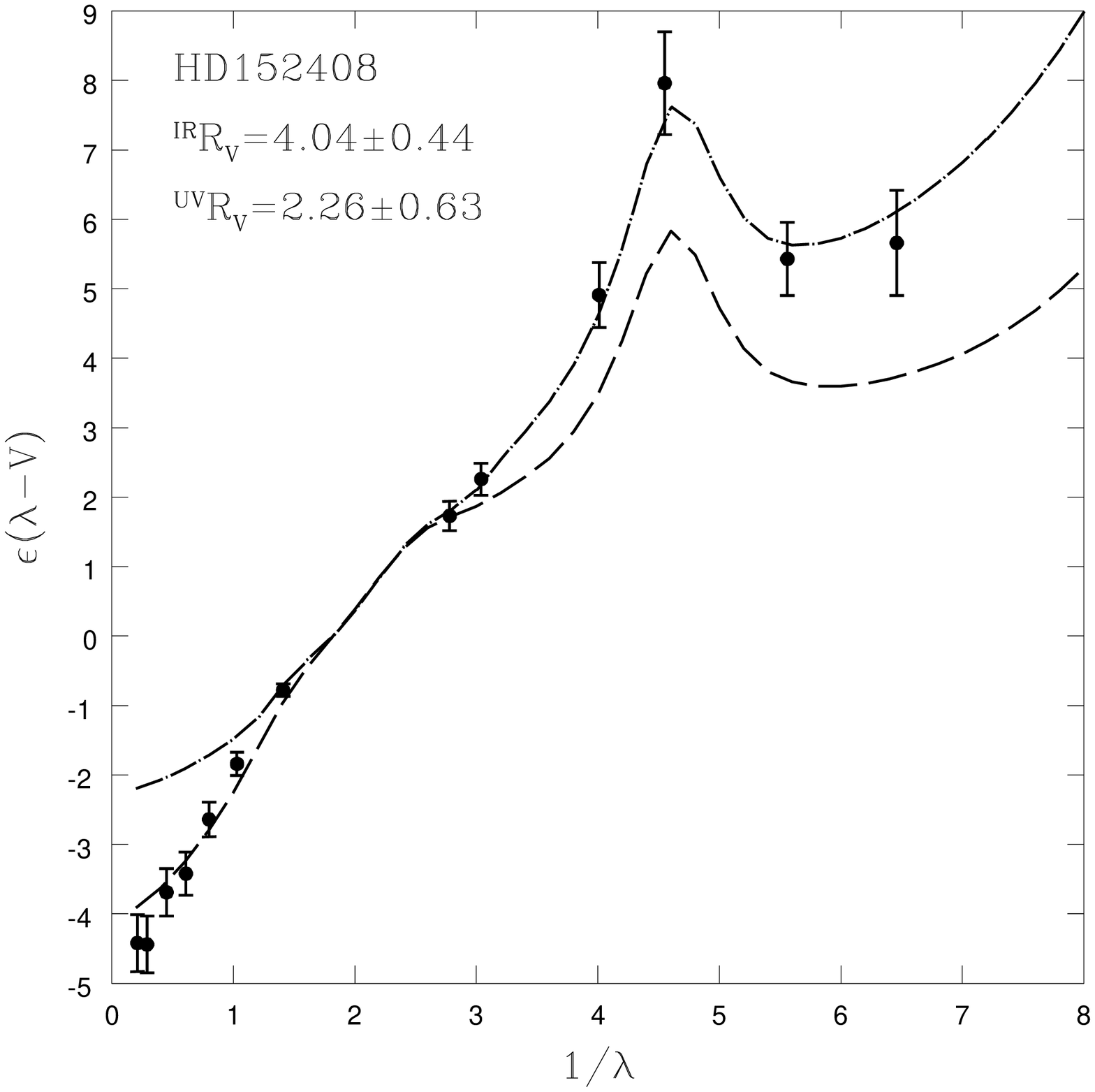}
\includegraphics[height=10cm,width=8cm]{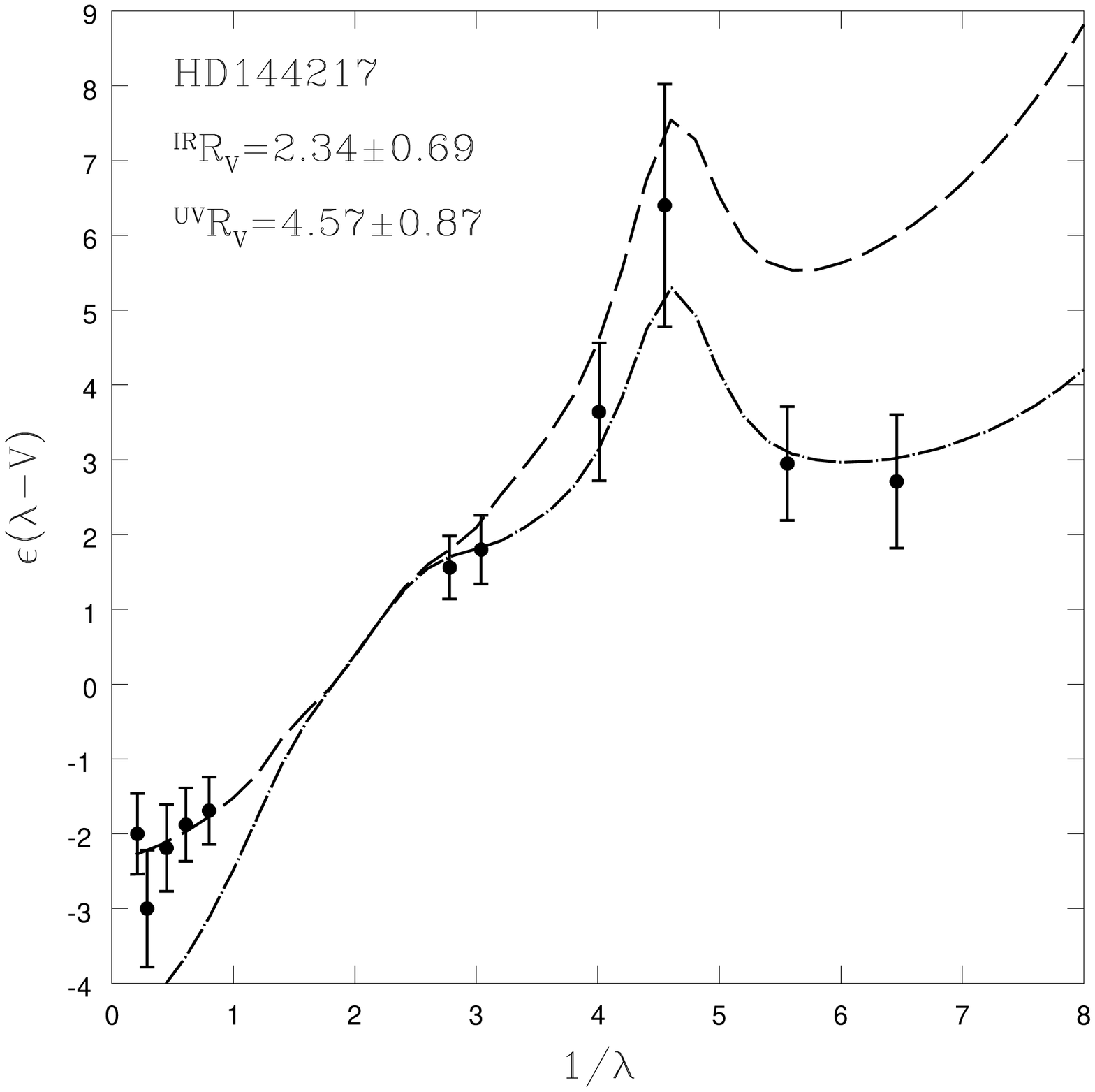}
\caption{Examples of extinction curves for which two different $R_V$
values are necessary to obtain a good fit to the entire extinction curve.}
\label{fig4}
\end{figure}

Figure \ref{fig3} shows the comparison between the $R_V$ values
obtained from optical/IR and UV extinction data. There is a large scatter
around the 1-to-1 relationship; however, only the points with the
error bars shown are those which deviate from this relation significantly 
($|\delta| \geq 2.0$). These 25 lines of sight need two 
different values of $R_V$ to match the optical/IR and UV part of the
extinction curve properly. Therefore, in these cases we cannot
universally use the CCM law to represent the entire extinction
curve. Six cases out of 25 have the $^{\rm{UV}}\!R_V$ higher than the 
$^{\rm{IR}}\!R_V$ one, and the other 19 have the $^{\rm{UV}}\!R_V$
lower than the $^{\rm{IR}}\!R_V$ one. Figure \ref{fig4} shows two
examples of such anomalous extinction curves.

\section{Conclusion}

We use a $\chi^2$ minimization method to compute the $R_V$ values for
a sample of 436 lines of sight. We exclude 20 cases because of their
negative $R_V$ and analyze the final sample of 416 lines of sight. We
derive our $R_V$ values assuming CCM law. This law aims at
reproducing the entire extinction curve by using only one parameter:
$R_V$. We analyze the goodness of the CCM fit for all lines of sight
using the $\chi^2/dof$ statistic and test the universality of the CCM
law by computing $R_V$ values separately for the optical/IR and UV
part of the extinction curve. We find that for 69\% of our original
sample, the CCM law is able to fit well both the optical/IR and UV data.
We divide the remaining 31\% of cases into two groups, according to two
main peculiarities: a) the optical/IR or/and UV extinction data points
cannot be fitted by the CCM law (25\% of the entire sample); b) $R_V$
values are significantly different for the two spectral regions:
optical/IR and UV (6\% of the entire sample). Unless caused by faulty
data, peculiar extinction curves result from unusual properties of
dust grains. Therefore, theoretical modeling of these extinction
curves (e.g., Mishchenko \cite{Mishchenko89}; Saija et al. \cite{Saija01}; 
Weingartner \& Draine \cite{Weingartner01}) may help us to understand
the processes which modify the properties of interstellar grains.

\end{document}